\documentclass{aastex61}
\usepackage{natbib}
\usepackage[version=3]{mhchem} 
\usepackage{graphicx}

\newcommand\aastex{AAS\TeX}

\received{13/08/2018}
\accepted{24/10/2018, ApJL}

\shorttitle{\aastex\ Early assembling of the Solar System}
\shortauthors{Pignatale et al.}

\begin{document}

\title{Making the planetary material diversity during the early assembling of the Solar System}

\correspondingauthor{Francesco C. Pignatale}
\email{pignatale@ipgp.fr}

\author[0000-0003-0902-7421]{Francesco C. Pignatale}
\affil{Institut de Physique du Globe de Paris (IPGP) \\
1 rue Jussieu, 75005, Paris, France}
\affil{Mus\'eum national d’Histoire naturelle, UMR 7590, CP52 \\
57 rue Cuvier, 75005, Paris, France}

\author{S\'ebastien Charnoz}
\affil{Institut de Physique du Globe de Paris (IPGP) \\
1 rue Jussieu, 75005, Paris, France}

\author{Marc Chaussidon}
\affil{Institut de Physique du Globe de Paris (IPGP) \\
1 rue Jussieu, 75005, Paris, France}

\author{Emmanuel Jacquet}
\affil{Mus\'eum national d’Histoire naturelle, UMR 7590, CP52 \\
57 rue Cuvier, 75005, Paris, France}


\begin{abstract}
VERSION: ArXiv    DATE: \today

Chondritic  meteorites, the building blocks of terrestrial planets, are made of an out-of-equilibrium assemblage of solids formed at high and low temperatures, either in our Solar system or previous generations of stars. This was considered for decades to result from large scale transport processes in the Sun's isolated accretion disk. However, mounting evidences suggest that refractory inclusions in chondrites formed contemporaneously with the disk building.

Here we numerically investigate, using a 1D model and several physical and chemical processes,  the formation  and transport of rocky materials during the collapse of the Sun's parent cloud and the consequent Solar Nebula assembling. 

The interplay between the cloud collapse, the dynamics of gas and dust, vaporization, recondensation and thermal processing of different species in the disk, results in a local mixing of solids with different thermal histories.  Moreover, our results also explains the overabundance of refractory materials far from the Sun and their short formation timescales, during the first tens of kyr of the Sun, corresponding to class 0-I, opening new windows into the origin of the compositional diversity of chondrites.

\end{abstract}




\section{Introduction} 
\label{intro}

Chondritic meteorites are fragments of planetesimals formed in the Solar Nebula within the first 4 Myr of Solar System history (the time 0  being the formation of the highly refractory inclusions, Ca-, Al- rich, CAIs) \citep{2015GMS...212....1C}. They are made of a mixture of materials that formed at high and low temperatures \citep{2003TrGeo...1..143S}. This peculiar assemblage has always been puzzling and is one key ingredient in the composition of terrestrial planets. A striking paradox is that carbonaceous chondrites, which presumably accreted in the outer solar system at temperatures lower than 150 K \citep{2011GeCoA..75.6912W}, are the richest in  CAIs formed at temperatures higher than 1600 K, presumably close to the Sun \citep{2003TrGeo...1..143S}. The chemical composition of CAIs links them to precursors condensed from the solar gas \citep{2006M&PS...41...83R}. 

Absolute Pb-Pb ages show that CAIs in carbonaceous chondrites  were formed over a brief interval (time, $t<160$~kyr) at the very beginning of the Solar System \citep{2012Sci...338..651C} from refractory precursors that, according to short-lived $^{26}Al$ chronologies, could have condensed over an even shorter period of 10-20 kyr  \citep{2008E&PSL.272..353J,2011ApJ...735L..37L,2014E&PSL.390..318M}. Such a timescale is shorter than the assembling time of a protoplanetary disk from the infalling molecular cloud \citep{2011ARA&A..49...67W}. It seems thus unavoidable that the formation of the earliest solar system solids (CAIs at least) occurred concurrently with the building of the Solar Nebula, in agreement with observations of young stellar objects showing that dust grows and becomes largely crystalline as soon as the star begins to form  \citep{2011ARA&A..49...67W,2012ApJ...756..168C}. Since increasing observational and theoretical evidence shows that disks start off compact \citep{2006ApJ...640L..67D,2011A&A...526L...8J,2016ApJ...830L...8H,2017ApJ...851...83C}  most of the infalling material  should be located initially close to the protostar and undergoes thermal processing before  incorporation into planetary material, like chondrites or planetesimals. 

To understand the origin of this material, we have developed a 1D simulation to track the formation of the disk and the dust thermal evolution during the collapse of the parent cloud.

\section{Methods and models}
\label{methods}

The code aims at tracking the 1D evolution of a  protoplanetary disks, with different gaseous and solid species. We use a simplified treatment of the gas and dust dynamics and chemistry (nevertheless including a variety of relevant coupled processes) in order to unveil key physical processes associated to the coupling of transport and thermal processing in a connected cloud-disk system. This is the reason the infalling envelope is described as a spherical isothermal shell collapsing while conserving angular momentum, in order to make the calculations computationally  tractable (see e.g. \citet{2005A&A...442..703H}). Although more detailed models are available for the cloud collapse, e.g. \citet{2016ApJ...830L...8H},  and numerous are  the studies investigating the disk chemistry \citep{2013ChRv..113.9016H}, our approach represents the first attempt to  implement in one single simulation several dynamical and chemical aspects  of cloud collapse and disk formation and evolution.

The code is a classical alpha-disk model built on the model of \citet{2005A&A...442..703H} that considers cloud collapse and radiative and viscous heating.  Our 1D disk is described by a logarithmic grid (100 cells, between 0.01 and 1000 au). We include gravitational instability, dead zone  in the form of a parameterized description \citep{2008ApJ...689..532T}. We use a layered accretion model of \citet{2010ApJ...713.1143Z} with a low-$\alpha$ midplane and high-$\alpha$ surface layers, dust transport through advection and diffusion, and growth/fragmentation using the method desribed by \citet{2016A&A...594A.105D} and \citet{2012A&A...539A.148B}. The opacity is calculated  using the opacity table from  \citet{2016A&A...590A..60B} and procedures described in \citet{2000A&A...358..651H} and in \citet{2003A&A...410..611S}. Starting from the work of   \citet{2012M&PS...47...99Y} on refractory dust, we now include multiple chemical species allowing material condensation and sublimation using a simple chemical model.

The infall of the parent cloud is described as a source term and here below we  describe  the main aspects. The cloud collapses onto the protostar-disk system with a constant accretion rate \citep{1977ApJ...214..488S}: 
\begin{equation}
 \dot{M}= 0.975\frac{C^{3}_{cd}}{G} ,\
\label{eq1}
\end{equation}
where $G$ is the gravitational constant and $C_{cd}$ is the isothermal sound speed\footnote{ $C_{cd}^{2}=k_{b}T_{cd}/\mu m_{p}$ where $k_{b}$ is the Boltzmann constant, $T_{cd}$ is the cloud temperature, $\mu=2.2$ is the molecular weight of the gas in terms of the proton mass, $m_{p}$.} in the cloud. Assuming that the angular momentum of the cloud material is conserved during the infall, the surface density accreted below a threshold radius, called ``centrifugal radius'', is \citep{2005A&A...442..703H,2012M&PS...47...99Y}:
\begin{equation}
\dot{\sigma}(r,t)= \frac{\dot{M}}{8\pi R^{2}_{c}} \Bigg( \frac{r}{R_{c}(t)} \Bigg)^{-3/2}  \Bigg[  1 - \bigg(\frac{r}{R_{c}(t)} \Bigg)^{1/2} \Bigg]^{-1/2} ,\
\label{eq2}
\end{equation}
where the centrifugal radius $R_c(t)$ is the radius under which the cloud collapses onto the disk. $R_c(t)$ is initially so small that all infalling material falls directly onto the protostar. As the star mass increases, $R_c(t)$  increases and the cloud material feeds progressively the disk. $R_c(t)$  is determined as the place where the specific angular momentum of the infalling material is equal to the specific angular momentum of the keplerian disk.  Thus, deeper cloud envelopes fall close to the protostar while exterior cloud envelopes fall far from the protostar. The expression for the centrifugal radius is
\begin{equation}
R_c(t)= 53 \Bigg( \frac{\omega_{cd}}{10^{-14} s^{-1}}  \Bigg)^{2}   \Bigg( \frac{T_{cd}}{10~K}  \Bigg) ^{-4}     \Bigg( \frac{M(t)}{1 M_{\odot}}  \Bigg) ^{3}   ,\
\label{eq23}
\end{equation}
where $\omega_{cd}$  is the constant angular velocity of the cloud, $T_{cd}$ is the temperature of the cloud and $M(t)$ is the total mass of the protostar+disk system at time $t$. 

Concerning the chemistry,   a simplified approach is used where simple phase transitions are considered. Using a thermochemical code to compute in real time the exact chemical evolution of the different species in the simulation is very challenging. Indeed, at each time step and  location the local gas and dust composition changes. This implies the implementation and tracking of at least several dozens of compounds if only a “short” list of elements were considered (see for example the shortened list of 60 species that can be stable in protoplanetary disk conditions  derived in  \citet{2011MNRAS.414.2386P} using the fifteen most abundant Sun's elements  \citep{2009ARA&A..47..481A}).

We, thus, introduce  a simplified chemical model (but nevertheless physically motivated, as the temperatures of condensation/processing of numerous species are well established \citep{1976ARA&A..14...81B,2006mess.book..253E}), in order to preserve the main trends of the chemistry.  In our simulations, the bulk chemical composition of the parent cloud is assumed to be solar  \citep{2009ARA&A..47..481A,2011MNRAS.414.2386P} with all the major rock-forming elements (Si, Mg, Al, Ca, Fe) hosted in interstellar unprocessed solid phases. We allow the infalling interstellar materials to be either vaporised, “processed” (i.e. heated at temperatures $T > 900$~K  enough to thermally alter the dust and destroy presolar grains but below vaporisation) or unaltered upon its arrival in the disk. 

The model tracks the formation and transport in the disk of gaseous species and different solid species with their own vaporisation temperatures  (See Table~\ref{table1}: (i) condensed refractory phases, which are the potential precursors of CAIs present in meteorites, named hereafter, for sake of simplicity, CAIs, (ii) condensed metal named hereafter metallic iron, (iii) condensed silicates, (iv) thermally processed and unprocessed silicate dust, (v) thermally processed and unprocessed iron dust, (vi) water ice, and (vii) CO ice. Molecular hydrogen is then the main gas species.

Chondrules formation is neglected,  as we essentially model their precursors. However, chondrule formation is not expected to affect large-scale compositional fractionations in the disk \citep{2015MNRAS.452.4054G}. Heating from accretion shock  is also neglected because of the low temperatures involved \citep{2008A&A...491..663D}.

\begin{table*}
\begin{tabular}{|c|c|c|c|c|}
\hline 
species & T (K) & become & T  (K) & become \\ 
\hline 
unprocessed refractories &  $>1650$  & refractory(g) & &  \\ 
\hline 
refractory(g)  &   $1500 <T<1650$ & CAIs$^{a}$ & $>1650$   &   refractory(g)  \\ 
\hline 
refractory(g) & $<1500$ & condensed silicates$^{a,b}$ &  &  \\ 
\hline 
unprocessed refractories &  $<1500$  & considered as silicates$^{b}$ & &  \\ 
\hline 
unprocessed silicates & $>900$  & processed silicates & $>1500$  & silicate(g) \\ 
\hline 
silicates(g) & $<1500$ & condensed silicates$^{a}$ & $>1500$ & silicate(g) \\ 
\hline 
unprocessed iron & $>650$ & processed iron & $>1550$ & iron(g) \\ 
\hline 
iron(g) & $<1550$ & metallic iron & $>1550$ & iron(g) \\ 
\hline 
metallic iron & $< 650$ & processed iron$^{c}$ &  &  \\ 
\hline 
water ice  & $> 150 $ &  water  vapour & $< 150 $   & water  ice   \\ 
\hline 
CO ice & $> 25 $ &  CO(g) & $< 25 $ & CO ice  \\ 
\hline
\ce{H2}(g) &  &   &  &  \\ 
\hline 
\end{tabular} 
\caption{Simple chemistry rules implemented in the code. We assume that the kinetics timescales of evaporation, condensation and processing are instantaneous or with a kinetics close to the  timesteps used in the calculations, $\delta$t~$10^{-2}$~yr. (a) In our calculations condensed CAIs and condensed silicates do not equilibrate with the surrounding environment as evidenced by actual chondrites. (b) This allows us to distinguish between the most refractory Ca-Al-bearing phases (such as hibonite and melilite) that condense at $1500< T\rm{(K)}<1650$ (those we are more interested in) and the other less refractory Ca and Al bearing silicates (like the fassaite phase) that condense at  $T<1500$~K together with the main silicates  \citep{2011MNRAS.414.2386P}. As a consequence our estimation of ``CAIs'' production can be considered  conservative. (c) Metallic iron is defined processed at $T<650$~K as it can undergo  chemical alterations such as sulfidation  and oxidation \citep{1976ARA&A..14...81B,2006mess.book..253E}.}
\label{table1}
\end{table*}

\section{Results and Discussion}
\label{results}

\begin{figure}
{\includegraphics[width=1.0\columnwidth]{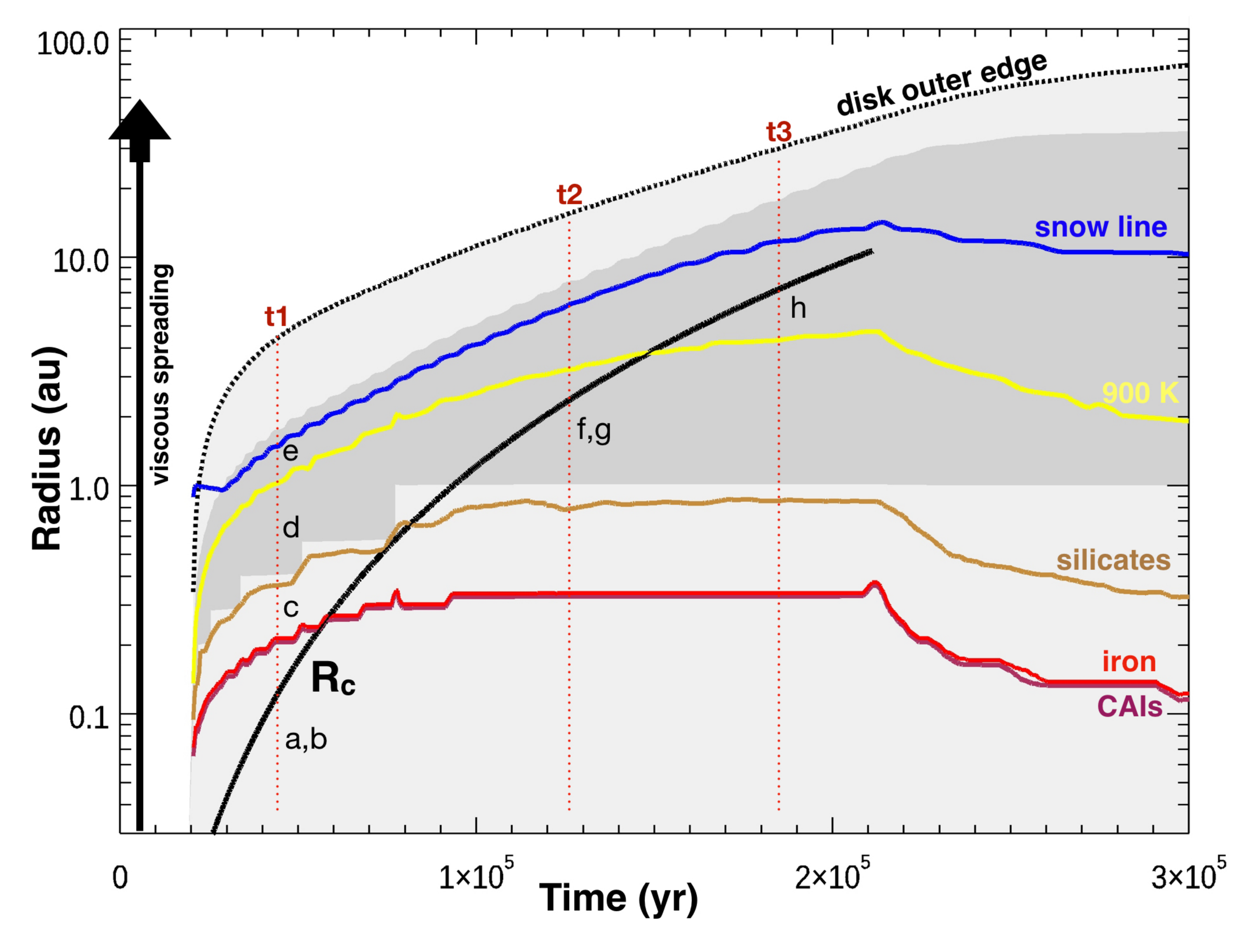}}
\caption{Time evolution of the centrifugal radius and the locations of disk outer edge, different condensation fronts, and dead zone (dark grey). Processes taking place at three snapshots are indicated. At $t\sim50$~kyr (t1), $R_{c}(t)$ is within the condensation fronts of the most refractory species (CAIs and iron) so that all the dust injected with the gas (a) is vaporised (b). At the same time, further out in the disk, condensation of CAIs and iron (c) silicates (d) takes place from a gas which is spreading outwards. Iron condensed  and transported with the spreading gas can be processed further out (e). At $t\sim120$~kyr (t2), $R_{c}$(t) has reached  a region in the disk where the temperature is below 1500 K. Presolar refractory material and silicates can be injected into the disk without being vaporised (f) but they can experience thermal processing (g). At $t\sim180$~kyr (t3), parent cloud material is injected in the disk likely retaining its primordial texture as the temperature is too low for thermal processing (h).} \label{fig1}
\end{figure}

Figure~\ref{fig1} shows the time evolution of the centrifugal radius and the locations of disk outer edge, different condensation fronts, and dead zone (dark grey). With our initial conditions ($T_{cd}=15$~K, $\Omega_{cd}=10^{-14}$ rad/s, $M_{0,\star}=0.02M_{\odot}$, $T_{\star}=4000$~K, $R_{\star}=3R_{\odot}$, $L_{\star}=constant$, $M_{tot}=1 M_{\odot}$,  $\alpha_{active}=10^{-2}$, $\alpha_{dead}=10^{-5}$, $v_{frag}=10~{\rm ms^{-1}}$), the cloud infall and disk’s assemblage lasts $\sim220$ kyr. While being fed, the disks also expands due to viscous spreading \citep{2005A&A...442..703H,2012M&PS...47...99Y}. After $t\sim220$ kyr the material starts flowing mostly inwards as the disk accretes onto the proto-Sun. During the infall, a dead zone (DZ), a region of low diffusivity \citep{2008ApJ...679L.131T,2008ApJ...689..532T}, develops between $\sim0.1$ and $\sim10$~AU.

As the dust is incorporated into the disk, its  composition changes because of heating of minerals with different sublimation temperatures.  Our simulations show that the final distribution of minerals is controlled by the interplay between two key parameters: (i) the centrifugal radius, $R_{c}$(t), i.e. the distance (increasing with time) within which the parent cloud material is injected and (ii) the location of the different condensation fronts and properties of the different mineral species, parameters never  fully considered previously. Because of angular momentum constraints, $R_{c}$(t) is initially close to the forming Sun fed from the parent cloud's innermost regions \citep{2016MNRAS.461.2257K}. Then $R_{c}$(t) increases with time due to the arrival of originally more distant cloud materials with higher specific angular momentum. Under our initial conditions, $R_{c}$(t) increases from 0 to $\sim10$ AU when the infall ends (Fig.~\ref{fig1}). The location of the condensation fronts of the different solid species considered  depends on the disk temperature profile  which is, itself, controlled by viscous and radiative heating \citep{Charnoz2018}. For a given species, if $R_{c}(t)$ is inside the condensation front of that species, then the dust will be totally vaporised upon its injection in the disk. However, the disk spreads outward concurrently, so that gas injected at high temperature cools down while moving outward, allowing its condensation at larger heliocentric distance. Moreover, dust injected at later stage, in a lower temperature environment, can be just processed without being vaporised, or survive thermally unaltered. These processes are sketched in Fig.~\ref{fig1}.

\begin{figure}
{\includegraphics[width=1.0\columnwidth]{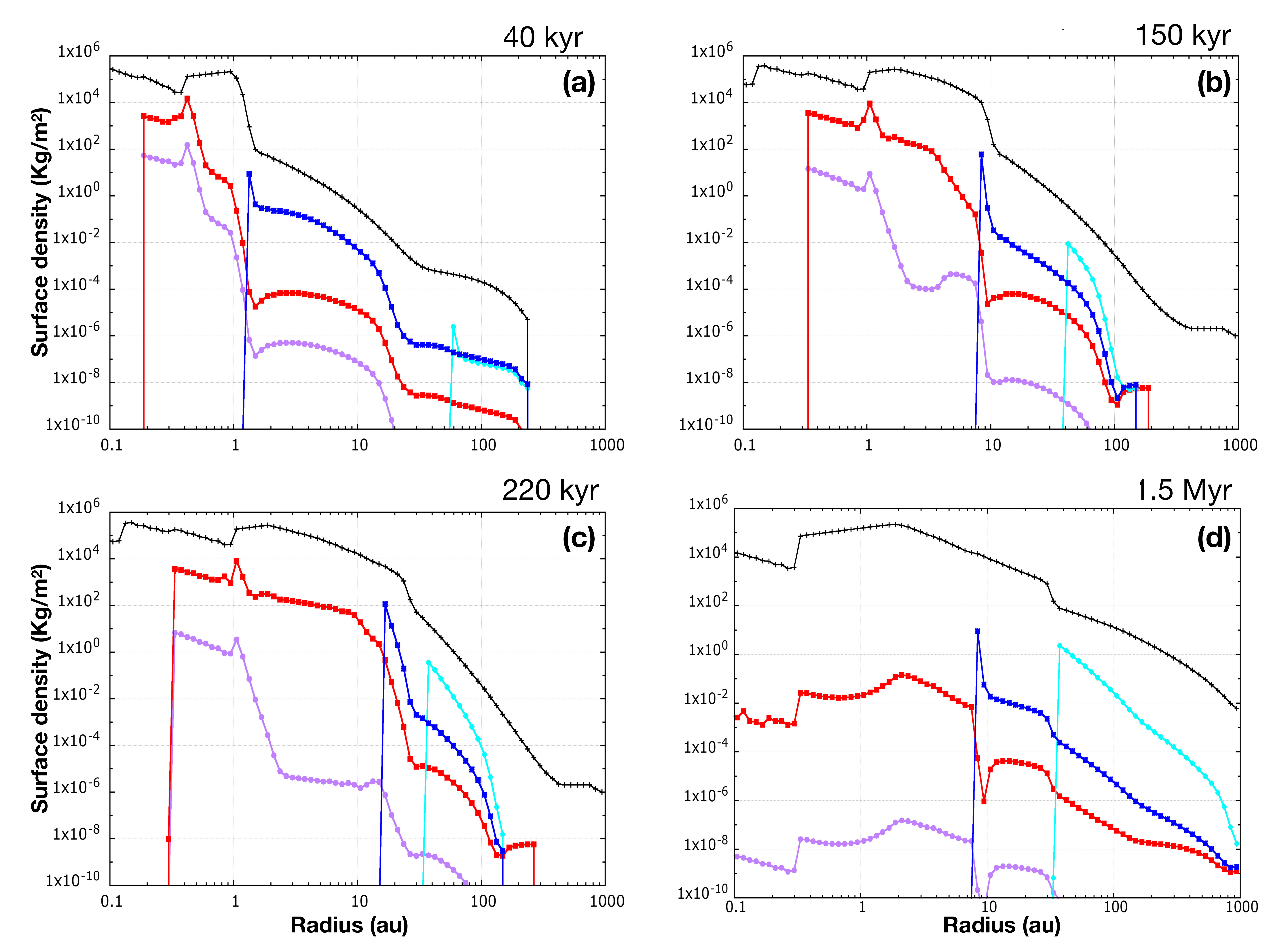}}
\caption{Surface density profiles at four different timescales.  \ce{H2(g)} black,  \ce{H2O}(ice) blue,  \ce{CO}(ice) cyan, CAIs purple,  and other rocky material in red. It can be  seen how the disk increases its mass and expands in the early stages (a,b,c, with c the end of collapse) and then looks like an accretion disk at later stage (d).}
\label{fig2}
\end{figure}

\begin{figure}
{\includegraphics[width=1.0\columnwidth]{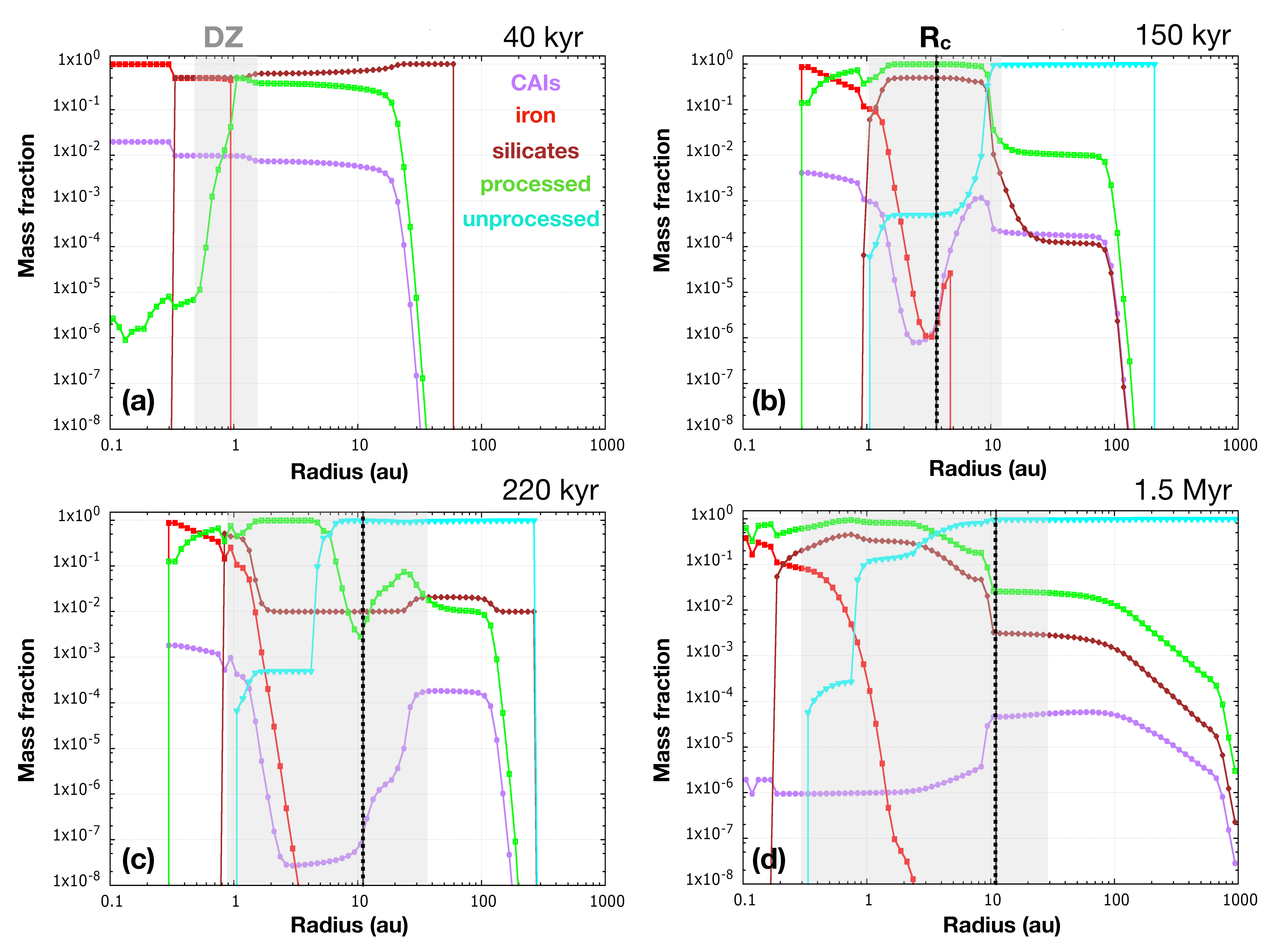}}
\caption{Mass fraction of various types of rocky material in the disk as a function of radius and time.  The different types of solids considered are CAIs (purple), metallic iron (red), condensed silicates (brown), all processed dust (green), all unprocessed dust (cyan).  The four panels show the calculated time integrated dust distribution at four snapshots, using the modelled disk evolution described in Fig.1. The light grey zone shows the extension of the dead zone. The black dotted line shows the location of the centrifugal radius $R_{c}(t)$. As $R_{c}(t)$ moves outward, material is injected in the disk at different temperatures, thus experiencing different thermal processing. Similary to \citet{2012M&PS...47...99Y} the relative decrease of CAIs starting at $t\sim150$~kyr results from a dilution due to the injection of parent could material directly from the infalling cloud,  that in our case fall into the disk’s Dead Zone. } 
\label{fig3}
\end{figure}

Figures ~\ref{fig2} and~\ref{fig3} show four time-snapshots  of the surface density profiles of gas, dust and ices and the radial distribution of the  mass fraction of the considered rocky material. Our results show that at $t\sim40$ kyr (Fig.~\ref{fig3}(a)) the disk is only populated by condensed dust (CAIs, iron and silicates) and only one kind of processed dust: condensed metallic iron that could experience low temperature chemical alteration ($T<650$ K, see Table~\ref{table1}). Later, as $R_c(t)$ moves outward, parent cloud dust can be preserved if a given mineral is injected outside its condensation front. At $t\sim150$~kyr, $R_c(t$) penetrates the DZ (Fig.~\ref{fig3}(b)) where dust and gas can be stored because of low turbulence. This lowers dramatically the concentration of refractory condensates within the DZ. At the end of the infall, most of the DZ has been fed by parent cloud material, leading to a DZ poor in refractory condensates and a CAIs-enriched outer disk beyond the DZ (Fig.~\ref{fig3}(c)). Inside the inner half of the DZ, material can be further thermally processed as the temperature could reach $T\sim900$ K while the outer half of the DZ contains mostly thermally unaltered dust. Our simulations also show that this heterogeneous dust distribution will be qualitatively preserved over the lifetime of the disk (Fig.~\ref{fig3}(d)). However, inward the DZ inner edge ($R<0.3$ AU), the disk becomes CAIs-poor because of strong inward accretion of gas that removes solids. 

This distribution predicted for the various dust components in the disk (Fig.~\ref{fig3}(d)) shows striking similarities with the composition of the three main families of chondritic meteorites (enstatite, ordinary, carbonaceous chondrites) as defined by their fractions of CAIs, metal, chondrules and matrix \citep{2003TrGeo...1..143S}. To do this exercise of comparison, we consider to first order the unprocessed dust as a proxy for the fine-grained matrix containing different kinds of circumstellar grains \citep{2003E&PSL.209..259N} and GEMS (glass with embedded metal and sulfides) which, for a fraction of them, are likely surviving interstellar grains \citep{2011GeCoA..75.5336K}. Within 1-10 AU, dust is dominated by disk-born silicates (processed or condensed), iron (processed or condensed, inside 2 AU essentially) and poor in CAIs and pristine interstellar dust. This is reminiscent of (matrix-poor, chondrule-rich) non-carbonaceous chondrites, i.e. ordinary chondrites and enstatite chondrites (the latter likely formed well inside the water snow line).  Further away in the disk, the dust is dominated by unprocessed ISM dust that was injected late when $R_c(t) > 5$~AU, processed dust and condensed silicates. This region, though beyond the snow-line and thus having water ice-bearing dust, has the highest fraction of CAIs of the whole disk. This bears similarities to carbonaceous chondrites. Note the presence, in addition to CAIs, of condensed silicates formed from the same parent gas reservoir. These silicates could then correspond to the Amoeboid Olivine Aggregates (AOAs) of carbonaceous chondrites that are known to have strong links with CAIs (from their oxygen isotopic composition, see \citet{2004ChEG...64..185K}). 

\begin{figure}
{\includegraphics[width=0.5\columnwidth]{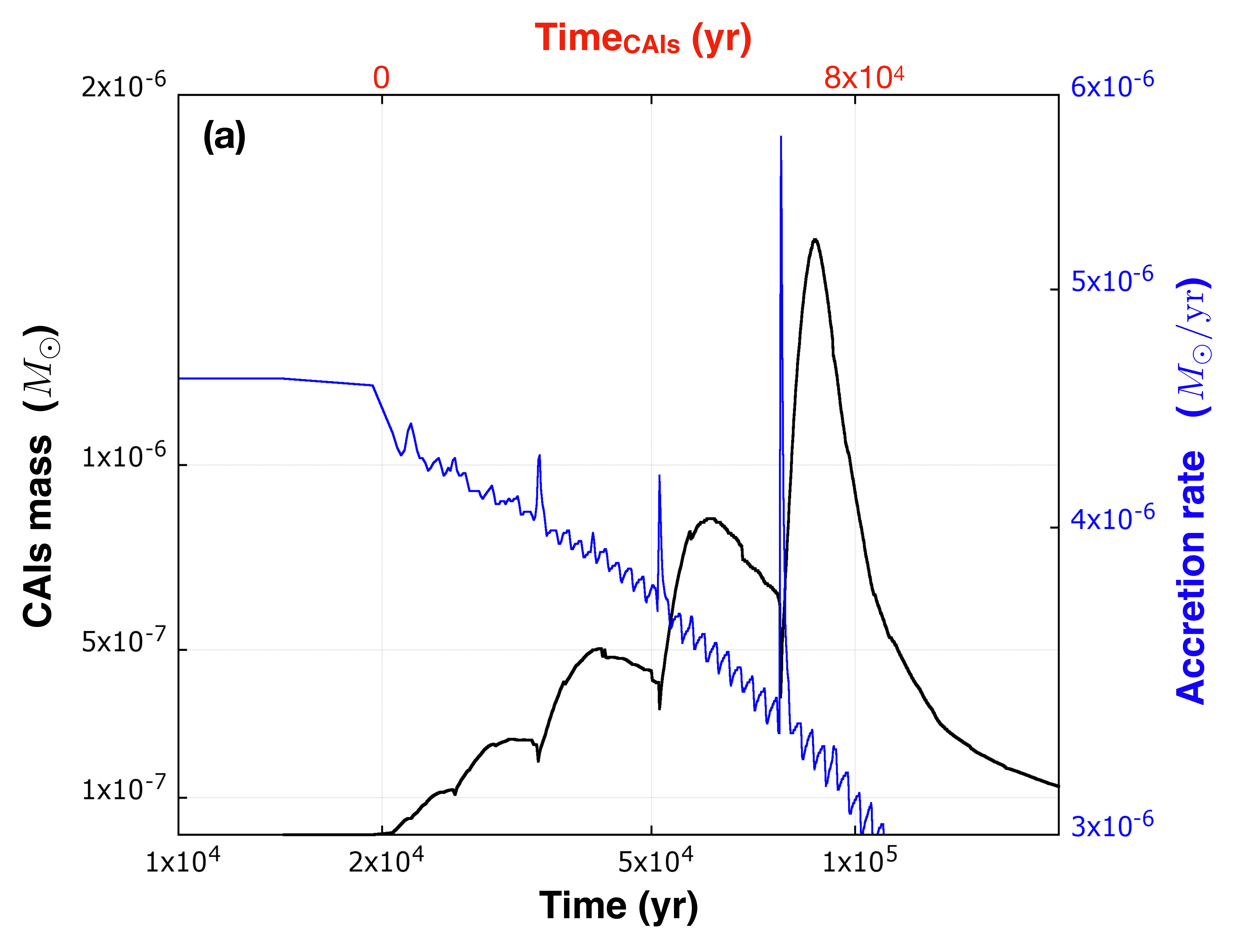}}
{\includegraphics[width=0.5\columnwidth]{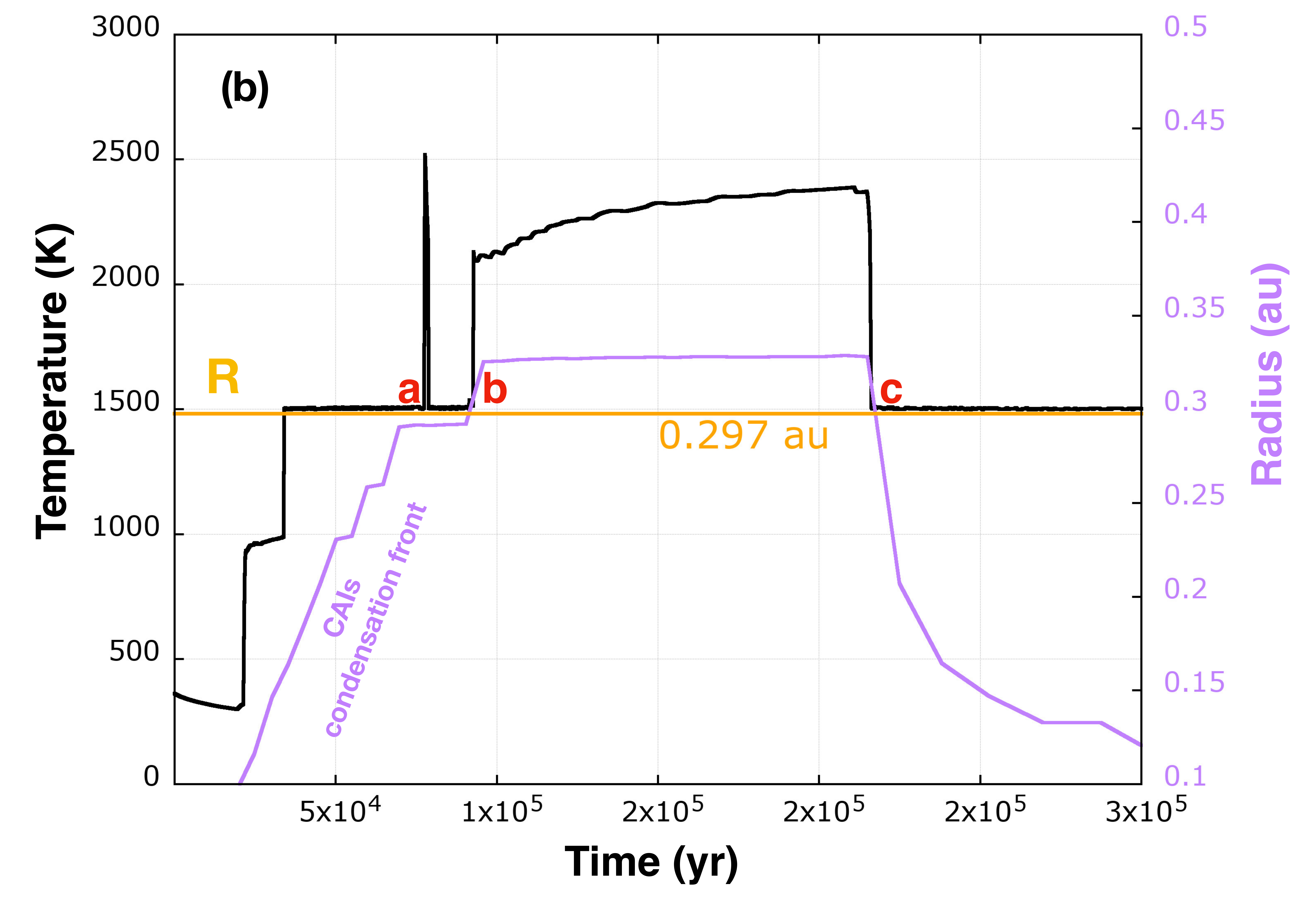}}
\caption{(a): Time evolution of the CAIs mass in the disk (black line, left y-axis) and of the accretion rates (blue line, right y-axis). The total CAIs mass in the disk is in solar mass and accretion rates in solar mass per year.  The zero of the age scale $Time_{CAIs}$ (upper horizontal axis) corresponds to the start of CAIs condensation in the simulations. Abrupt steps in the CAIs mass and accretion rates are observed between 0 and $t\sim70$~kyr ($Time_{CAIs}$). In the simulation, four transient outbursts occur. These events are like ``FU Orionis'' events \citep{2009ApJ...701..620Z}, and are followed by material evaporation and re-condensation. They are caused by mass loading of the DZ due to the gas inflow from the disk and the parent cloud. These episodes are transient because the enhanced viscosity induces a fast spreading of material. The bursts in production of CAIs occur after episodes of sudden accretion, the strongest one being the one at $Time_{CAIs}\sim70$~kyr. (b): Time evolution of the temperature in the disk (black line left y-axis) at a selected radius ($R=0.297$~AU) and the time evolving location of the CAIs condensation front (purple line right y-axis). The condensation front of CAIs is the location of the $T=1650$~K isotherm (note that after $t\sim220$~kyr when the flow of gas reverses towards the star in this region, this front inverses its displacement too and goes closer to the star). The temperature  profile evidences the presence of an outburst (an abrupt pulse with $T>2000$~K). If these pulses in temperature take place in a region outward the condensation front of CAIs, already formed CAIs can experience thermal alteration or partial evaporation (it is the case of the point marked with ``a'' in the figure).  All the CAIs that are located at a given radius are vaporised when the CAI condensation front crosses this radius (marked ``b''). CAIs can then re-condense as the condensation front moves back toward the inner disk regions (marked ``c''). Note that temperatures as high as those required to evaporate the precursors of FUN CAIs ($T\sim2200$~K \citep{2013GeCoA.123..368M} can be reached in these outbursts.
} 
\label{fig4}
\end{figure}

Interestingly, formation of CAIs is found to be a transient event. In Fig~\ref{fig4}(a) we report the time evolution of the CAIs mass in the disk and the time evolution of the accretion rates. In Fig.~\ref{fig4}(b) we show the time evolution of the temperature in a given location of the disk and the time evolving location of the CAIs condensation front. 

Massive production of CAIs happens, in the simulation, from $t\sim40$ kyr to $t\sim70$ kyr, in very good agreement with Pb--Pb and 
$^{26}\rm {Al}$--$^{26}{\rm Mg}$ chronologies of CAIs \citep{2012Sci...338..651C,2014E&PSL.390..318M} (Fig.~\ref{fig4}(a)). This corresponds to a  short time window, when interstellar solids are injected in the disk at temperatures high enough for them to be fully evaporated (i.e. when $R_c(t)$ is inward the condensation radius of CAIs, see Fig.\ref{fig1}), so that upon spreading outward, the gas can condense into CAIs.  This time is shorter by a factor of a few than that found by \citet{2012M&PS...47...99Y} because of their lower condensation temperature for refractories.

Still, the production of CAIs appears to occur by steps, linked to accretion bursts onto the star following mass loading and instabilities of the DZ \citep{2009ApJ...701..620Z,2014MNRAS.445.2800O} (Fig.~\ref{fig4}(a)). This may help to explain part of the diversity of CAI populations across chondrite groups \citep{2014mcp..book..139M} despite overall derivation from the same reservoir (the inner disk in our model).  Possibly, some of the reheating episodes experienced by many CAIs \citep{2006M&PS...41...83R,2017E&PSL.472..277S}  may be linked to such bursts of accretion, re-heating,  partial vaporisation condensations (as suggested in Fig.~\ref{fig4}(a,b)) although many of those episodes were likely of much shorter duration.  Our results set the formation of the main bulk of CAIs precursors found in chondrites as early as the formation of the disk itself (class 0 to I).

\section{Conclusions}
\label{conclusions}

In this work we investigated the effects of the cloud collapse and disk building in forming and  distributing  different solids species in the disk. We find  that  coupling  dust transport and thermal  processing during the cloud collapse and disk formation naturally results in  local mixing of materials with different thermal histories, in qualitative agreement with the three main chondritic families. Our model also explains the overabundance of refractory materials far from the Sun and their short formation timescales, during the first tens of kyr of the Sun. Our finding can reconcile  all these long-existing cosmochemical questions within one single picture  and place the time 0 of the Solar System during the class 0 to I of the Sun's formation history.

Some additional processes  (e.g. disk winds acting at the disk surface) may redistribute material at large distance, and a more refined chemistry can give more precise predictions. However, this simple model unveils new ways, never described before,  to interpret the distribution of chondritic material in our Solar System, that may have kept the imprint of the could infall. This works implies that it is now critical to study the protoplanetary disk, not in isolation, but in the context of the early environment of the Sun which has left fingerprints in the diversity of our Solar System material.

\acknowledgments
The authors wish to thank the anonymous referee for the detailed and useful comments which largely improved the manuscript. The authors wish to acknowledge the financial support of ANR-15-CE31-0004-1 (ANR CRADLE) and of the UnivEarthS Labex programme at Sorbonne Paris Cit\`e (ANR-10-LABX-0023 and ANR-11-IDEX-0005-02)

%

\vspace{5mm}

\end{document}